\documentclass{rspublic}
\usepackage{amsmath}
\usepackage{amsfonts}
\usepackage{amssymb}
\usepackage[Symbol]{upgreek}
\usepackage[nointegrals]{wasysym}
\usepackage{mathrsfs}
\usepackage{graphicx}
\newcommand{\real}{{\mathbb R}}
\newcommand{\bfu}{{\bf u}}
\newcommand{\bfx}{{\bf x}}

\newcommand{\Upo}{{\Sigma}}
\newcommand{\bdy}{{B}}
\begin{document}
\title[Navier-Stokes Equations in Bounded Domain]{\large Viscous flow regimes in a square. Part 1. Lid-driven cavity}
\author[F. Lam]{F. Lam}
%
%
\label{firstpage}
\maketitle
\begin{abstract}{Planar Navier-Stokes Equations; Vorticity; Turbulence; Diffusion; Viscosity; Randomness; Enstrophy; Palinstrophy}

In the present paper, we examine the viscous flow evolution in a square cavity. Coupled with the stream function, the initial-boundary value problem of the vorticity is numerically solved by a method of iteration. The only boundary condition is the wall velocity which in turn defines the Dirichlet value for the stream function. We assert that the corner singularity in the cavity flow is in fact a theoretical artefact. By adopting suitably regulated lid velocity, excellent comparison with experiments is found because of the converged palinstrophy field. Our calculations demonstrate that asymmetrical shears initiate the formation of the vortices with counter-rotating strained cores, while the consecutive re-birth and the subsequent disintegration of the developed eddies proliferate the shears into smaller scales. This production process is particularly intense in close proximity to the solid surfaces. 
\end{abstract}
\section{Introduction}\label{intro}
The most important aspect of flow motion in the presence of hydraulically smooth solid boundaries is to quantify the vorticity production at the surfaces. The problem can be rephrased as the study of the convection and diffusion of the vorticity field in connection with energy dissipation. As the fluid viscosity becomes small or Reynolds number increases, experimental observation shows that the production processes are actively operational in a turbulent shear-laden region in the solid vicinity, known as a boundary layer. In this viscous near-wall region, physics dictates that momentum transfer must occur, as there is an increase in the wall shear stress, giving rise to enhanced energy consumption. Thus understanding of the near-wall fine-scale vorticity structures is of importance. To avoid the controversy due to approximations, such as an analytical singularity in the boundary layer theory, it is best to work with the full set of the Navier-Stokes equations, so as to identify the genuine dynamics effects. Surprisingly, our endeavour to underpin a self-consistent and efficient numerical method for $2D$ turbulence is a matter of continuing debate. To bypass the main difficulties in assigning the wall vorticity, many of the previous works on rectangles concentrate on periodic boundary conditions which are non-physical. Naturally, any Navier-Stokes flow, developed from arbitrary initial condition, must be inhomogeneous, anisotropic and highly unsteady. We believe that certain fundamental aspects on the governing differential equations have long been overlooked, leading to inconsistency among the proposed numerical schemes. This view is particularly justifiable for the vorticity-stream function formulation. For an introduction to planar vorticity dynamics and its computation, see Gresho (1991).

In the present note, we discuss the issues related to the flow evolution as well as the numerical simulation inside the square cavity of unit side length.
\subsection*{Equations of motion}
\begin{figure}[ht] \centering
  {\includegraphics[keepaspectratio,height=7cm,width=7cm]{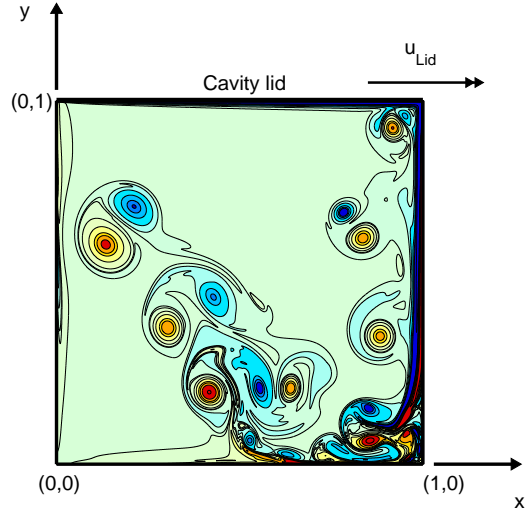}} 
 \caption{The layout of the unit square cavity. The lid velocity moves to the right as indicated. We consider $u_{\rm{Lid}}$ as a function of $x$ so that the overall vorticity dynamics can be better modelled, particularly at small viscosity.} \label{cvlayout} 
\end{figure}

The equations governing the planar incompressible viscous flows are
\begin{equation} \label{ns}
	\Delta_t \bfu = (\partial_t   - \nu \Delta )\bfu = - (\bfu. \nabla )\bfu  - {\rho}^{-1} \nabla p, \;\;\; \mbox{and} \;\;\; \nabla.\bfu = 0,
\end{equation}
where the velocity is denoted by $\bfu=(u,v)$, the pressure by $p$. We use notation $\bfx=(x,y)$ for the space variables. All the other symbols have their usual meanings in fluid dynamics. The flow domain ${\Upo}$ is $0  \leq x \leq 1 \cup 0 \leq y \leq 1$ with boundary $\bdy$ (see figure~\ref{cvlayout}). In every problem of fluid motion, there is an initial state: the initial velocity is denoted by
\begin{equation} \label{ns-ic}
 \bfu(\bfx,0)=\bfu_0(\bfx)\;\;\; \in \;\; C^{\infty}(\Upo).
\end{equation}
We assume that $\bfu_0$ is solenoidal and is given. The boundary condition is 
\begin{equation} \label{ns-bc}
 \bfu_0(\bfx,t)=\bfu_{\bdy} \;\;\; \forall \bfx \;\; \mbox{on} \;\; \bdy,
\end{equation}
where $\bfu_{\bdy}$ is prescribed. In applications, the usual no-slip condition refers to $\bfu_{\bdy} \equiv 0$.

The continuity equation in (\ref{ns}) implies that $u \rd y {-} v \rd x$
is always an exact differential of the independent variable $\bfx$ at any given time $t$.  There exists a function $\psi(\bfx;t)=\psi(\bfu(\bfx);t)$ whose definition reads
\begin{equation} \label{psidef}
	\rd \psi = u \rd y - v \rd x,
\end{equation}
where the components of the solenoidal field are given by
\begin{equation} \label{vdef}
u(\bfx;t) = \partial_y \psi(\bfx;t) \;\;\;\;\;\; \mbox{and} \;\;\;\;\;\; 
v(\bfx;t) = -\partial_x \psi(\bfx;t).
\end{equation}
As the velocity on solid surfaces is given in practical application, we determine the $\psi$-value on the boundary by integrating (\ref{psidef}) 
\begin{equation} \label{psibc}
	\psi_{\bdy}(\bfx;t) = \int_{\bdy} \big( \: u(\bfx;t) \; \rd y - v(\bfx;t) \; \rd x \:\big) + \psi_0 = f(\bfx;t).
\end{equation}
This is nothing more than a re-statement of the velocity boundary conditions.
The vorticity, $\zeta = \partial_x v - \partial_y u$,
is a solenoidal quantity which describes angular momentum of fluid particles. Taking the curl operation on (\ref{ns}), we obtain the vorticity-stream function formulation 
\begin{equation} \label{vort}
\Delta_t  \zeta  =- \partial_y \psi \; \partial_x \zeta + \partial_x \psi \; \partial_y \zeta, \;\;\;\;\;\;  
\Delta \psi = - \zeta,
\end{equation}
where the Poisson equation follows the vorticity definition. The initial vorticity data are denoted by
\begin{equation} \label{vt-ic}
	\zeta(\bfx,0)=\zeta_0(\bfx)= \nabla\times \bfu_0(\bfx) \;\;\; \forall \bfx \in \Upo.
\end{equation}
There are no boundary data for the vorticity as they are part of the solutions.
It is well-known that the vorticity in $\real^2$ satisfies the maximum principle:
\begin{equation} \label{max-vort}
	\max_{\bfx \in \real^2, \; t \geq 0 } \big|\: \zeta \: \big| \;{\leq}\; \max_{\bfx \in \real^2 } \big| \: \zeta_0 \: \big|,
\end{equation}
see, for instance, John (1982). This {\it a priori} bound asserts the global regularity of the vorticity equation. But the pressure field is not known to possess similar bounds and, specifically, its well-posedness cannot be established without the knowledge of the vorticity. Indeed the pressure serves as an auxiliary variable in the description of fluid motion. Because of fluid's propensity to deform and to shear, fluid dynamics is all about the variations of fluid particles' angular momenta. The scalar pressure does not contribute to the mechanical torque. 
Nevertheless, the pressure is completely determined once the vorticity and the velocity are known. More precisely, the pressure is well-defined as long as the vorticity and its derivatives are locally integrable (i.e. $\zeta, \zeta_x, \zeta_y \in L^1_{loc}$). Taking the divergence operation on (\ref{ns}), we obtain the Poisson equation
\begin{equation} \label{modp}
\Delta p/\rho = 2 \: \big( u_x \: v_y - u_y\: v_x \big). 
\end{equation}
The Neumann boundary conditions, $\partial_x p$ and $\partial_y p$, are found from (\ref{ns}). 

By the vector identity $\nabla{\times}(\nabla{\times}A)=\nabla(\nabla.A)-\Delta A$, we have
\begin{equation} \label{uv1}
	\Delta \bfu = - \nabla {\times} \zeta,
\end{equation}
if the continuity is satisfied. Then the momentum equation in (\ref{ns}) can be re-written as
\begin{equation} \label{modns}
\partial_t u + \nu \partial_y \zeta = v \zeta - \partial_x \chi, \;\;\;\;\;\;
\partial_t v - \nu \partial_x \zeta = -u \zeta - \partial_y \chi,
\end{equation}
where $\chi=p/\rho+(u^2+v^2)/2$ is the Bernoulli-Euler pressure, and can be found from the Poisson equation 
\begin{equation} \label{modp1}
\Delta \chi = \zeta^2 - u \partial_y \zeta + v \partial_x \zeta. 
\end{equation}
Equations (\ref{modns}) provide the Neumann boundary data on the cavity walls,
\begin{equation} \label{pwall}
	\partial_y \chi \: \big|_{x=0,1} \;\;\; \mbox{and} \;\;\; \partial_x \chi \: \big|_{y=0,1}.
\end{equation}
The surface integral of the right-hand side of (\ref{modp1}) is 
\begin{equation} \label{compati}
	\int_{\Upo} \Big(\zeta^2 - u \zeta_x + v \zeta_y \Big)\rd \bfx = 
		- \int_{(y=0,1)} u \zeta \:\rd \bar{x}  + \int_{(x=0,1)} v \zeta \:\rd \bar{y}
\end{equation}
from integration by parts. The last two formulas are line integrals. (As the lid has a non-zero velocity, only the integral along $y{=}1$ remains while the other three vanish.) Assuming the boundary data are time-invariant, we evaluate the quantity 
\begin{equation*}
\int_{(x=0,1)} \partial_x \chi \: \rd\bar{y} + \int_{(y=0,1)} \partial_y \chi \:\rd\bar{x}
\end{equation*}
using (\ref{modns}). We find that the sum of the terms involving viscosity $\nu$ and the corner vorticity vanishes due to cancellation, and that the remaining two parts with integrand containing $(\bfu, \zeta)$ exactly equal to the line integrals in (\ref{compati}). Thus, the Neumann problem of Poisson equation (\ref{modp1}) (and (\ref{modp}) for $p$) is compatible with boundary data (\ref{pwall}). This solvability constraint substantiates the view that there exist no {\it a priori} pressure boundary conditions.

The dimensions of the momentum equation, and of the continuity in (\ref{ns}) are $[\rm{m} \: \rm{s}^{-2}]$ and $[\rm{s}^{-1}]$ respectively. Physical laws must be independent of the units used to measure a particular system. We consider the equations of motion (\ref{ns}) as inherently dimensionless with respect to the SI standard length, time and mass, 
namely $1$ metre ($\rm{m}$), $1$ second ($\rm{s}$), and $1$ kilogram ($\rm{kg}$). These standards are precisely defined. Then the viscosity becomes dimension-independent relative to the unit viscosity $\nu_0{=}[\rm{m}^2 \:\rm{s}^{-1}]{=}1$ (thus a derived base unit). By the same token, the SI base units imply the existence of the unit dynamic viscosity $\mu_0{=}[\rm{kg} \: \rm{m}^{-1} \: \rm{s}^{-1}]{=}1$. In this respect, the vorticity and pressure equations are unitary and do not depend on the (arbitrary) choice of a case-dependent characteristic velocity, or of a length scale.
\section{Lid-driven cavity} \label{ldc}
Roughly speaking, we do not have precise information on the inertia behaviour of the lid at the starting instant. This is even true in experimental investigations (see, for instance, Guermond {\it et al}. 2002). As far as the computations are concerned, the crucial specification is the lid's velocity which is assumed to be an acceptable representation of the physics. Following a common practice, the initial velocity (\ref{ns-ic}) is theorised by a step function:
\begin{equation} \label{ics}
	\bfu_0(\bfx)=u_0(0 \leq x \leq 1,y=1) = \begin{cases}
	\; 0, & t < 0, \\
	\; u_{\rm{Lid}} , & t \geq 0.
	\end{cases}
\end{equation}
Should an admissible $u$-distribution be given, the vorticity dynamics evolves according to this specified boundary data. In this respect, the initial vorticity is somehow loosely defined and will affect the flow in a short time interval immediately after the start, $t \sim 0^+$. As our interest lies in the transient stage away from the starting phase, we merely need to specify an approximate initial shear which will be superseded by the vorticity field compatible with the lid velocity. We examine a few options for $u_{\rm{Lid}}$ and hence $\zeta_0$. Their suitability is assessed in the light of comparison with experiment. 

We assume that the two upper corners are perfectly sealed in laboratory experiments. If a tiny gap at each of the corners is permissible, fluid is effectively allowed to enter and move out of the cavity. For instance, if we move along the side wall $x=0$ into the cavity, the local jump in the $u$-velocity, 
$\partial_x u |_{x=0, y \rightarrow 1} \neq 0$.
As the flow is incompressible, the amount of the flow-in must be balanced by that of the flow-out at the right-hand corner. The net fluid inside the cavity has to remain unchanged over time; the law of mass conservation ought to be interpreted by the integral form of the continuity equation $\int_{\Upo} (\nabla. \bfu) \rd \bfx =0$.
\subsection*{Irregular lid velocity}
At $t=0$, the upper lid is impulsively set into a horizontal motion at constant speed of unity, as given in (\ref{ics}). The no-slip applies to the side and lower walls. To avoid the double-valued velocity at each of the upper corners, we specify the lid velocity as
\begin{equation} \label{bcs}
u_{\rm{Lid}} = \begin{cases}
	\; 1, & 0< x < 1,  \\
	\; 1/2, & x=0,\;x=1.
	\end{cases}
\end{equation}
Since the jump at $x=0$ (or $x=1$) occurs from $0$ to $1$, the last condition, the arithmetic mean at the jump discontinuity, can be justified by a standard Fourier analysis where the two side walls are periodically extended in the $y$-direction. 

In the immediate vicinity of the lid, the continuity shows that the normal velocity $v$ may be an arbitrary (bounded) function of $x$, if the lid velocity is independent of $x$. This fact in turn suggests that the initial vorticity, $\zeta_0 \sim \partial_x v$, may be specified without much constraint except the trivial shear-free case. 

The simplest initial data may be approximated by an impulsive motion in terms of the generalised function:
\begin{equation} \label{imps2}
\zeta_0(x,y) = - \delta(y-1),
\end{equation}
which implies a vorticity sheet of vanishing thickness in the limit $\nu \rightarrow 0$. Alternatively, as the impulsively-started flow resembles the initial motion of linear heat equation, we may describe the initial vorticity by a Stokes layer
\begin{equation} \label{imps}
\zeta_0(x,1) = - \frac{1}{\sqrt{4 \pi \nu \Delta t}},
\end{equation}
where $\Delta t$ equals the marching time step. The numerics involved in these two scenarios, at vanishing viscosity, may present a challenge for practical computations.
\subsection*{Regularised problem}
In experiment, a moving lid at constant speed is the easiest way to set-up cavity flows. However, the uniform speed does not mean that the viscous wall layer has to be uniform across the whole lid, due partly to the blockage by the side walls. At the upper right corner, there exists no physical space for the lid fluid to move to the right, and the corner must constitute a stagnation point. At the left, if a portion of fluid moves with the lid to the right, the available fluid must come from the side wall as it would be drawn upward. On the other hand, the fluid particles residing near $x=0, y \approx 1$ must be stationary, in order to satisfy the no-slip. Then the local continuity implies that no fluid may come from the wall, nullifying any non-zero corner flow. At some distances away from the corners, it is plausible that the local flow may separate, inducing confined circulating vortices.

Our specification of the corner velocity (\ref{bcs}) will certainly misrepresent the physics, in all likelihood; the corner singularity must be a theoretical artefact. It is the velocity discontinuity which will corrupt the whole flow-field. For computational purposes, it is entirely justified for us to restore to regulated lid velocity. 

The following bell-shaped velocity has been proposed
\begin{equation} \label{bell}
	u_{\rm{Lid}}= 16 \: x^2(1-x)^2,
\end{equation}
where the unity attains only at the mid-location $x{=}0.5$. It is easy to see that the initial vorticity may be set as
\begin{equation} \label{bellzt0}
	\zeta_0(x,y)= -32 \:y \: ( 1 - 6 x + 6 x^2),
\end{equation}
if the incompressibility hypothesis is respected near the lid. Then the initial motion is assumed as an impulsive type and starts with the lid-vorticity $\zeta_0(x,1)$. 

\begin{figure}[ht] \centering
  {\includegraphics[keepaspectratio,height=7cm,width=12cm]{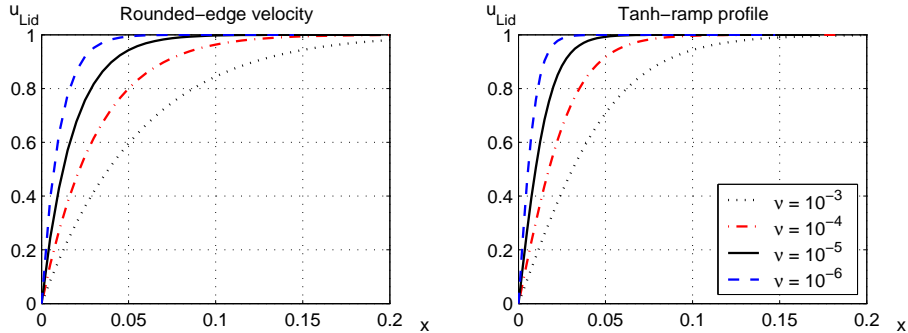}} 
 \caption{Regulated lid velocity: rounded edges (\ref{round}) and tanh-ramp profile (\ref{ramp}). The $\nu$-dependence implies that a corner singularity is only possible at vanishing viscosity. } \label{profv} 
\end{figure}

An improved starting flow, which perhaps better simulates experimental set-up, is given by
\begin{equation} \label{round}
	 u_{\rm{Lid}} =   
\begin{cases}
	\: 1-(1-x)^\lambda, \: & x < 1/2, \\
	\:  1- x^\lambda, \: & x \geq 1/2,
	\end{cases}
\end{equation}
where the power of the monomial $\lambda = \pi/\nu^{1/4} $, and the fraction of constant velocity is extended. Furthermore, the corner regions can be delimited by a ramp profile 
\begin{equation} \label{ramp}
	 u_{\rm{Lid}} =   
\begin{cases}
	\: \tanh(\lambda \: x), & x < 1/2, \\
	\: \tanh\big(\lambda \: (1-x)\big), & x \geq 1/2,
	\end{cases}
\end{equation}
see figure~\ref{profv} for illustration. The starting vorticity may be estimated in line with the previous case. Clearly, the last two regularisations are reduced to a generalised $\delta$-vorticity in the limit of $\nu \rightarrow 0$. 
\section{Planar vorticity dynamics}
Dynamic system (\ref{vort}) is a non-linear diffusion equation for the vorticity. Because of the bound, such as (\ref{max-vort}), the equation is well-posed. Since we have no {\it a priori} knowledge of vorticity boundary values, we must determine them as part of the flow solutions. The most effective way is by means of iterative procedures. To see why we do not need any other boundary conditions apart from the $\bfu_{\bdy}$ data, we re-cast the vorticity evolution into an integro-differential equation:
\begin{equation} \label{vie}
\begin{split}
	\zeta(\bfx,t) = \zeta_s(\bfx,t) & + \int_0^t \!\! \int_{\Upo}  \big( \: \partial_{x'}\psi \: \partial_{y'} \zeta - \partial_{y'}\psi \:\partial_{x'} \zeta \:\big)(\bfx',s)\: H(\bfx,\bfx',t{-}s) \: \rd \bfx' \rd s \\
	\quad & \hspace{0.125cm} + \nu \int_0^t \!\! \int_0^1 \!\!  \big( H(\bfx,x',1,t{-}s) \: \eta_1 
	- H(\bfx,x',0,t{-}s) \: \eta_2 \big) \: \rd x' \rd s, \\
		\quad & \hspace{0.325cm} + \nu \int_0^t \!\! \int_0^1 \!\! \big( H(\bfx,1,y',t{-}s) \: \eta_3 - H(\bfx,0,y',t{-}s) \:\eta_4 \big) \: \rd y' \rd s,
\end{split}
\end{equation}
where $\eta_1(x,t),\eta_2(x,t),\eta_3(y,t)$, and $\eta_4(y,t)$ are the vorticity derivatives on $x=1$, $x=0$, $y=1$, and $y=0$ respectively. The kernel has a separable form 
\begin{equation} \label{hg}
	H(\bfx,\bfx',t)= \;N(x,x',t) \times N(y,y',t),
\end{equation}
where $N$ stands for the Neumann function,
\begin{equation*} 
\begin{split}
N(r,r',t) = & \Big[ 1 + 2 \sum_{n=1}^{\infty} \exp\big({-} n^2 \pi^2 \nu t \big) \cos(n \pi r) \cos(n \pi r') \Big] \\
\quad = & \frac{1}{\sqrt{4 \pi \nu t}} \sum_{n=- \infty}^{n = \infty} \! \Big[ \exp\Big({-} \frac{(r{-}r'{-}2n)^2}{4 \nu t}\Big) {+} \exp\Big({-} \frac{(r{+}r'{-}2n)^2}{4 \nu t}\Big) \Big].
\end{split}
\end{equation*}
The first term, $\zeta_s$, denotes the mollified initial vorticity. For instance, the initial lid impulse (\ref{imps2}) is mollified over time interval $\Delta t$ as 
\begin{equation} \label{icvt0}
	\zeta_s=\int_{\Upo} H(\bfx,\bfx',\Delta t) \zeta_0(\bfx')\rd \bfx' = - N(y,1,\Delta t),
\end{equation}
where we have made use of the well-known properties of the heat kernel. This result represents the vorticity field immediately after the start; $N \sim O(1/\sqrt{\nu \Delta t})$ which recovers the diffusion-dominated Stokes motion near the lid. Note that the starting shear (\ref{icvt0}) is bounded for any non-zero $\nu \Delta t>0$. Although the initial vorticity behaves like an impulse, its total strength has finite measures in space.
\subsection*{Auxiliary role of stream function $\psi$}
Use of the stream function guarantees the resulting velocity to be solenoidal. But the stream function does not have a precise meaning independent of the vorticity. 
Mathematically, streamlines are the (continuous) level-curves ${\rd x}/{u} = {\rd y} /{v}$ whose tangential velocities coincide with the direction of velocity $\bfu$, see \S 21 of Lagrange (1781). Lagrange dealt with irrotational potential flows, i.e., $\zeta \equiv 0$ so that $\Delta \psi=0$. His general theory implies that velocity $\bfu$ must be {\it non-vanishing} though this fact has not been explicitly stated. For rotational potential flows, see, for instance, Art. $154$ of Lamb (1975). Recall that the notion of viscous fluid dynamics had not been introduced in the eighteenth century. Should a level-curve of zero-mass-flow or a ``stagnation flow'' exist, the law of mass conservation is satisfied everywhere on the curve. The description of the stream function $\psi$ in these situations becomes superfluous as $\rd \psi \equiv 0$. In the presence of solid boundaries with the no-slip condition (say) in singly-connected regions, there are no streams flowing anywhere on the boundaries; any level-curve is now a trivial solution, $\psi|_{\bdy} \equiv 0$. In essence, {\it solid surfaces in viscous flows are not equivalent to streamlines}.   

For every given $\zeta$, the solution of the Poisson equation in (\ref{vort}) subject to (\ref{psibc}) is uniquely determined within a constant which does not cause any problem as we require only the derivatives. However, the converse, $\zeta=\zeta(\psi)$, amounts to a horse-before-the-cart approach to kinematics. Consider the extended stream function 
\begin{equation} \label{psi1}
	\tilde{\psi}(\bfx;t)=\psi(\bfx;t)+ \theta(\bfx;t).
\end{equation}
The linearity of the Laplacian asserts that the continuity and the Poisson relation hold for $\tilde{\psi}$ as well:
\begin{equation} \label{invort}
	\zeta=-\Delta \psi  = -\Delta \tilde{\psi},
\end{equation}
as long as $\partial_{xx} \theta=0$ or $\partial_{yy}\theta=0$ (at every given time $t$). Clearly $\theta$ may be any of the following functions or their linear superposition
\begin{equation*} 
 C_0(t),\;\;\; C_1(t) x, \;\;\; C_2(t) y, \;\;\; C_3(t)(x-x_0) (y-y_0),
\end{equation*}
where $C_0$ to $C_3$ are arbitrary non-zero finite constants, and the last term describes the stagnation flow centred at $(x_0,y_0)$. (In the case of simple geometries, method of images may be an effective way to offset incompatible boundary values.) This observation suggests that the stream functions alone cannot completely specify viscous incompressible flows. 

In numerical procedures, attempts to assign vorticity boundary data in terms of the stream function require full justification because any such specification is in general non-unique. 
Consider the example given on page 427 of Gresho (1991). Since $u_{\:\rm{Lid}}=\partial_y \psi$, and $v=-\partial_x \psi=0$, $\zeta=-\partial_{yy}\psi$ on the lid, these relations suggest that we may need to expand $\psi$ in a Taylor series on two grid rows just beneath the lid up to the third derivative $\psi_{yyy}$ which can then be eliminated from the two resulting expressions. In parallel to the standard derivations of finite difference formulas, we carry out the expansions and obtain
\begin{equation*}
	\zeta_{\: \rm{Lid}}=\zeta_{i,n+1} = \frac{3(A + B x - u_{\:\rm{Lid}})}{h} \:  + \frac{7 \psi_{i,n+1}- 8 \psi_{i,n}+ \psi_{i,n-1}}{2 h^2} + O(h)
\end{equation*}
where $h$ is the mesh size, and $A,B$ are arbitrary constants. They come from the extended stream function $\psi \rightarrow \psi + A y + B x y$ applied to one of the grid rows. By choosing similar extended $\psi$ on the remaining walls, we show that none of the vorticity boundary values is uniquely defined in terms of the stream function. In the technical literature, there are many variations of the $\psi$-expansion under exotic names, such as the Thom-Burggraf or the Wilkes-Pearson formula, or Jensen's method. 

\begin{figure}[ht] \centering
  {\includegraphics[keepaspectratio,height=6cm,width=15cm]{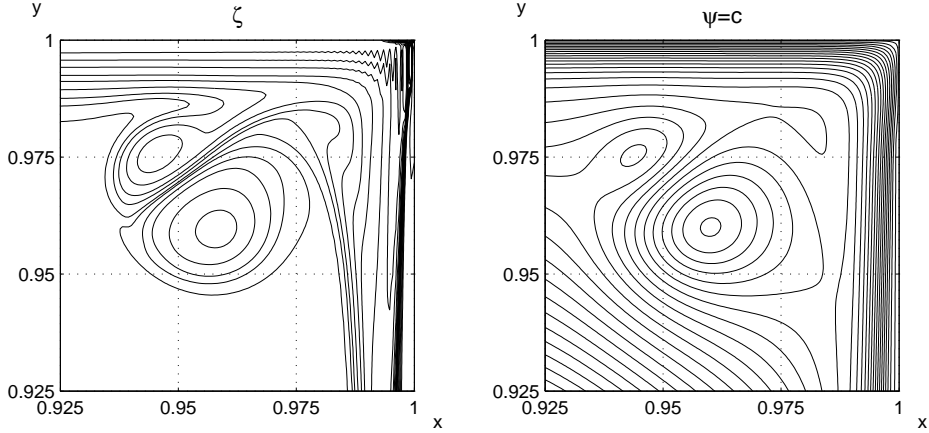}} 
 \caption{Numerical solution of ramp profile (\ref{ramp}) near the upper right corner at $t{=}2$. Run at $\nu=10^{-5}$, $\Delta t =5{\times}10^{-4}$, grid $n=1536$. The $\zeta$-contours are $\pm500$, $\pm150$, $\pm100$, $\pm75$, $\pm50$, $\pm35$, $\pm20$, $\pm15$, $\pm10,\pm5$ and $\pm1$. The $\psi$-field obtained from $\Delta^{-1}\zeta$ is highly regular at the current mesh which is however considered as inadequately-resolved for the vorticity field. It is reasonable to expect that the pure stream function formulation (\ref{sfeq}) would produce smooth, yet possibly spurious shears.} \label{rmpcnt} 
\end{figure}

By eliminating the vorticity in (\ref{vort}), an unsteady biharmonic equation governing $\psi$ can be derived
\begin{equation} \label{sfeq}
	\partial_t (\Delta \psi)  - \nu \Delta^2 \psi = \partial (\psi, \Delta \psi ) / \partial(x,y).
\end{equation}
In addition to the non-uniqueness issue in the foregoing discussion, solutions of this equation may also define ``moderated'' vorticity because the operator $\Delta \psi$ assumes certain regularity. In figure~\ref{rmpcnt}, we give an example which highlights a numeric aspect: the vorticity field is often far less smooth than the stream function. 
\subsection*{Solution of vorticity-stream function}
Consider the lid-driven flow where the lid velocity $u_{\rm{Lid}}$ is specified. Let us separate the solution (\ref{vort}) into $\psi = \psi^* + \bar\psi $, where 
\begin{equation} \label{psistar} 
	\Delta \psi^* =- \zeta,\;\;\;\;\;\; \psi^*_{\bdy}=0,
\end{equation}
and
\begin{equation*} 
	\Delta \bar\psi = 0,\;\;\;\;\;\; \bar\psi_{\bdy}(x,1)=u_{\rm{Lid}}(x).
\end{equation*}
By the method of separation of variables and Fourier series, the solution $\bar\psi$ can be expressed as  
\begin{equation*}
	{\bar\psi}(\bfx)= \sum_{m=1}^{\infty} A_m \: \sin(m \pi x) \sinh(m \pi y),
\end{equation*}
by virtue of the no-slip on the side and lower walls. As usual, the Fourier coefficients are given by
\begin{equation*}
	A_m=\frac{2}{\sinh(m \pi)} \; \int_0^1 u_{\rm{Lid}}(x)  \sin(m \pi x) \:\rd x.
\end{equation*}
Because $\partial_{xx}{\bar\psi} - \partial_{yy}{\bar\psi}=0$,
the solution ${\bar\psi}$ does not affect the vorticity. As we intend to iterate on the vorticity numerically, we need to solve the Poisson system (\ref{psistar}) for $\psi^*$. It follows that, at every iteration, the solution for $\psi$ is given by
\begin{equation*}
	\psi(\bfx)=\sum_{q=1}^{\infty}\sum_{r=1}^{\infty} B_{qr} \sin(q \pi x) \sin(r \pi y),
\end{equation*}
where
\begin{equation*}
	B_{qr}= \frac{4}{(q\pi)^2+(r \pi)^2}\int_0^1 \!\! \int_0^1 \zeta(x,y) \sin(q \pi x) \sin(r \pi y) \: \rd x \rd y.
\end{equation*}
This is an alternative to the full data (\ref{psibc}); the selective use of homogeneous Dirichlet data $\psi_{\bdy}=0$ may be more efficient in practical implementation of elliptic solvers. Our ideas may be generalised to arbitrary boundary data $f(\bfx)$ of $C^2$ domains.\footnote{There is a deep-rooted misconception that considers the Poisson boundary-value problem (\ref{vort}) as over-specified because two sets of boundary conditions have to be satisfied, namely Dirichlet and Neumann data (i.e., $	\psi (\bfx) = {\rm const} $, and $\partial \psi/\partial \vec{n}(\bfx) = {\rm const}$, where symbol $\vec{n}$ denotes the local outward normal). A host of numerical schemes then have been proposed to address the over-determinacy. Strictly speaking, the computational results from these formulations must be treated with caution as they can be misleading. The confusing state of affairs can be attributed to the view that solid surfaces in incompressible viscous flows identify with streamlines whose analytic exposition, i.e., the stream function, plays a unique and primary role in isolation.}
\subsection*{Solenoidal velocity from di-vorticity $\nabla{\times}\zeta$}
Instead of the stream function, the velocity can also be found from (\ref{uv1}) by inverting the Laplacian:
\begin{equation} \label{vv}
\begin{split}
u(\bfx) =& -\int_{\Upo} \;{\partial_{y'} G}(\bfx,\bfx') \;\zeta(\bfx') \rd \bfx' - \int_0^1  {\partial_{y'} G}\big|_{y'{=}1} \; u_{\rm{Lid}}(x') \:\rd x', \\ 
v(\bfx) =&  \int_{\Upo} \; {\partial_{x'} G}(\bfx,\bfx') \;\zeta(\bfx')  \rd \bfx',
\end{split}
\end{equation}
where Green's function $G$ satisfying homogeneous Dirichlet condition is given by
\begin{equation} \label{green}
	 G(\bfx,\bfx') =  2 \sum_{k=1}^{\infty} \: \sin(k \pi x) \sin(k \pi x')  
\begin{cases}
	\: \sinh(k \pi y') \sinh(k \pi (1{-}y)), & y' < y, \\
	\: \sinh(k \pi y) \sinh(k \pi (1{-}y')), & y' > y.
	\end{cases}
\end{equation}
The flow field is well-determined by solving the parabolic-elliptic system of (\ref{vort}) and (\ref{uv1}). Note that two Poisson equations must be inverted at every time step. 
\subsection*{Equivalence principle}
At the conclusion of iteration or convergence, the flow-field solutions given by 
\begin{equation}
	\zeta= \nabla {\times} \bfu,\;\;\;\;\;\; \bfu = (\partial_y \psi, \: -\partial_x \psi),\;\;\;\;\;\;   \Delta \bfu = - \nabla{\times}\zeta,
\end{equation}
must be self-consistent. This requirement defines an equivalence principle: numerical values evaluated from the vorticity-stream function formulation must agree with those from the di-vorticity, and vice versa. The principle may serve as a compatibility check in the numerical implementation.

The dynamics in (\ref{vie}) can be transformed into the canonical form containing quadratic non-linearity,
\begin{equation*}
	\zeta(\bfx,t) = \eta(\bfx,t) \;+ \int_0^t \!\! \int_{\Upo} K(\bfx,\bfx',t,s) \; \zeta^2(\bfx',s) \; \rd \bfx' \rd s,
\end{equation*}
where the kernel $K=K(G,N)$. In general, its solutions are of turbulence.
\subsection*{Numerical method}
\begin{figure}[t] \centering
  {\includegraphics[keepaspectratio,height=6cm,width=12cm]{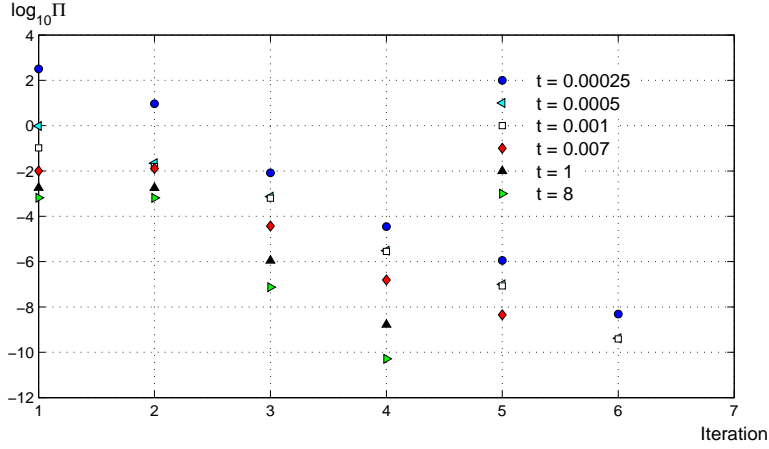}} 
 \caption{Convergence of ramp profile (\ref{ramp}) at $Re=1000$, $n=512$, and $\Delta t = 2.5{\times}10^{-4}$. } \label{rmpcv} 
\end{figure}

To seek the solutions of (\ref{vort}) or (\ref{vie}) by iteration, we subdivide the cavity into equally-spaced grids, where the vorticity equation is solved using any numerical schemes suitable for convection-diffusion equations. The grid or mesh is denoted by $n$ grid points or an $n^2$ mesh. Given the starting value $\zeta^0\equiv\zeta_0$, the implicit Euler scheme of time discretisation of the transport (\ref{vort}) may be symbolically written as
\begin{equation} \label{ztfd}
\begin{split}
 \big( 1 - \nu \Delta t D^2 \big) [\zeta]^{k+1}& = [\zeta]^{k} - \Delta t \: \big([\bfu]^{k+1}. D \big) [\zeta]^{k+1}, \\
D^2 [\psi]^{k+1}  & = - [\zeta]^{k+1},
\end{split}
\end{equation}
where $D$ stands for the centred finite difference operator, and it is understood that all flow variables cover all the grid points ($i,j$) within the square. The superscript $k$ denotes the iterations at the fixed local time. The system is closed with discretised $\psi_{\bdy}$ on the walls. The non-linear term at every start of the iteration loop may be approximated using the values of the previous time ($\zeta^k, \bfu^k$). This choice may speed up the computations by allowing larger $\Delta t$. In particular, we use a semi-implicit scheme for the non-linear term:
\begin{equation} \label{ztfd2}
- \frac{\Delta t}{2} \:\Big( \big([\bfu]^{k+1}. D \big) [\zeta]^{k+1}+ \big( [\bfu]^{k}. D \big) [\zeta]^{k}\Big).
\end{equation}
The penalty is that extra memory storage is required. To be efficient and flexible, the discretised system of the Poisson equation in (\ref{ztfd}) is solved by a multi-grid method. The resulting $n{\times}n$ vorticity matrix $\Pi$ is iterated until the convergence is achieved
\begin{equation} \label{cv}
	\delta \Pi=\sum_{ij} \Big| \;\Pi^{k+1}-\Pi^{k}\; \Big| < 10^{-8},
\end{equation}
where the sum is over all grid elements. An example is shown in figure~\ref{rmpcv}. To summarise, the solutions are obtained by the following algorithm:
\begin{enumerate}
	\item At the start $t=0$ (or $k=0$), the flow is specified by vorticity $\zeta^k$.
	\item March vorticity over $\Delta t$ to obtain $\zeta^{k+1}$. If norm $\delta\Pi$ satisfies (\ref{cv}), restart iteration using converged values as ``new initials''; otherwise, update elements: 
	$\; \zeta_{ij}^{k+1}\;\; \leftarrow \;\;\zeta_{ij}^{k+1} + \delta (\zeta_{ij}^{k+1}-\zeta_{ij}^{k})$. 
	\item Solve Poisson equation in (\ref{ztfd}) to determine interior $\bfu^{k+1}$. 
	\item Compute wall vorticity using $\zeta^{k+1}=\nabla{\times}\bfu^{k+1}$ (e.g. by finite difference).
	\item Renew non-linearity $(\bfu^{k+1}.\nabla)\zeta^{k+1}$. Loop back to step 2.  
\end{enumerate}
Step 4 is equivalent to specifying vorticity $\zeta_{\bdy}$ which is iterated in terms of the calculated velocity from Step 3 {\it and} the (fixed) boundary velocity $\bfu_{\bdy}$. This is why we may enforce the homogeneous Dirichlet condition $\psi_{\bdy}=0$ (\ref{psistar}) because, effectively, the process of fixing ($\zeta$,$\bfu$) does not involve $\psi_{\bdy}$. The velocity solutions are driven by the vorticity.

\subsection*{Integral properties}
No symmetry conditions have been assumed in the implementation of our numerical scheme. To monitor the progress of our computations, we examine energy,
\begin{equation} \label{e0}
	E(t)=\frac{1}{2} \int_{\Upo} \big( u^2 + v^2 \big)(\bfx,t) \;\rd \bfx,
\end{equation}
for fluid of unit density ($\rho=1$). 
From the momentum equation (\ref{ns}), we confirm the energy conservation
\begin{equation} \label{engy}
	\frac{1}{2} \int_{\Upo} \bfu^2 \; \rd \bfx + \nu \int_0^t \!\! \int_{\Upo} \zeta^2 \; \rd \bfx \rd \tau +\nu \int_0^t \!\! \int_0^1 u_{\rm{Lid}} \zeta(x,1) \; \rd x \rd \tau = \frac{1}{2} \int_{\Upo} \bfu_0^2 \; \rd \bfx,
\end{equation}
where the last term on the left is obtained provided $v_{\bdy}=0$. The accumulation of enstrophy over time is the quantity
\begin{equation} \label{q0}
	Q(t)=\int_0^t \int_{\Upo} \zeta^2 (\bfx,\tau) \;\rd \bfx \rd \tau = \int_0^t \Omega(\tau) \; \rd \tau.
\end{equation}
Thus energy dissipation by viscous action occurs at all times as long as there are non-zero shears anywhere in the flow field. The variation of the local quantity, $\Omega$, reflects the dynamical interaction of the vorticity. Furthermore, it may be interesting to inspect the circulation or the total vorticity flux inside the square, 
\begin{equation} \label{c0}
	\Gamma(t)=\int_{\Upo} \zeta(\bfx,t) \;\rd \bfx = \int_0^1 u_{\rm{Lid}}(x) \; \rd x,
\end{equation}
in view of Green's theorem. This quantity provides a useful and convenient check for numerical procedures in flows having some spatial symmetry. Lastly, the palinstrophy,
\begin{equation} \label{z0}
	Z(t)=\int_{\Upo} \big[ \:  \big( \partial_x \zeta \big)^2 + \big( \partial_y \zeta \big)^2 \: \big](\bfx,t) \;\rd \bfx,
\end{equation}
serves as a measure of the all-important non-linearity $(\bfu.\nabla)\zeta$, and has been taken as the benchmark for the quality of computational meshes. Given regular initial data, Heywood \& Rannacher (1982) show that it is the boundedness of di-vorticity that ensures numerical regularity, i.e., $	| \nabla \zeta |_{L^{\infty}(\forall \: \bfx \in \Upo, \; t > 0 )} < \infty$.
\section{Result and discussion}
Since we are interested in the transients, we first examine the impact of the initial-boundary data on the subsequent development. Once this issue has been understood, the detailed near-wall vorticity structures at small viscosity or large Reynolds number ($Re=\nu^{-1}$) are elucidated. In the cavity flows, an ``obvious'' choice of the characteristic velocity for Reynolds number is the maximum lid velocity which, however, does not fully specify the velocity {\it distributions} (cf. the lid profiles given in \S \ref{ldc}). The only exception is the case of a constant. 

\begin{figure}[t] \centering
  {\includegraphics[keepaspectratio,height=10cm,width=15cm]{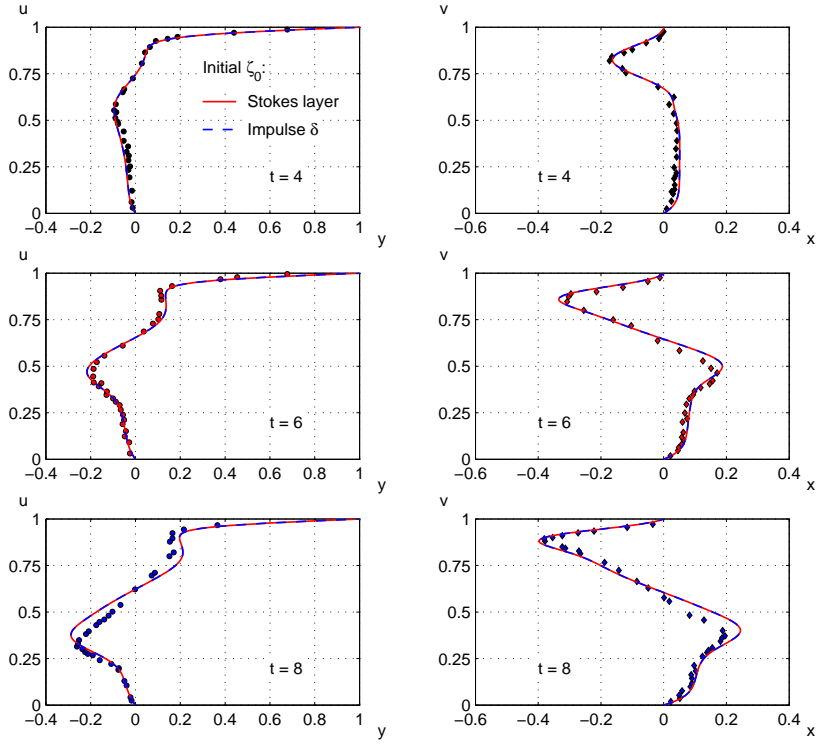}} 
 \caption{Comparison with the symmetry-plane measurements in a rectangular cavity with aspect ratio $1{:}1{:}2$ (Guermond {\it et al}. 2002). Assuming lid data (\ref{bcs}), the computations start from delta function (\ref{imps2}) or Stokes layer (\ref{imps}) ($Re{=}1000$, $n{=}256$). Symbols are the experimental points and solid lines the theory. The agreement appears impressive but is however {\it fictitious} as the theoretical flow contains irregular non-linearities, see figure~\ref{imphist}. Note that the computed velocity ($t>1$) is independent of the initial vorticity.} \label{impsing} 
\end{figure}

\begin{figure}[ht] \centering
  {\includegraphics[keepaspectratio,height=4.5cm,width=16cm]{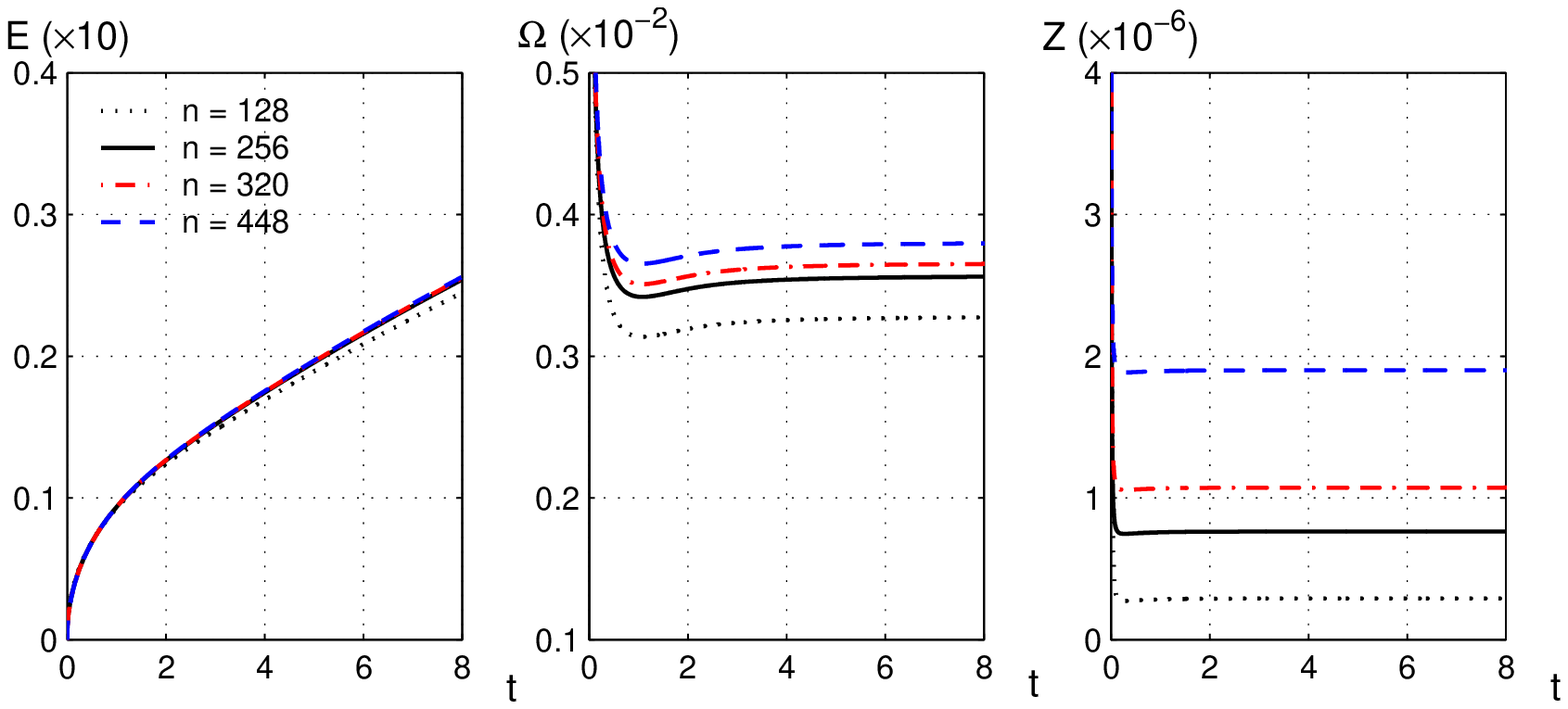}} 
 \caption{Lid velocity (\ref{bcs}) with starting Stokes layer (\ref{imps}) leads to mesh divergence in the enstrophy and palinstrophy though the energy appears to have converged. The magnitudes of these integral quantities are subject to the resolution of the singular corner flows. In numerical approximations, this means that the solutions are grid-dependent and non-unique. } \label{imphist} 
\end{figure}

The lid velocity distribution (\ref{bcs}) introduces essential singularities into the cavity. Shankar \& Deshpande (2000) argued that the singular corner flow is very localised so that cavity's velocity field is largely immune to any adverse effects. Such a view appears to have been confirmed in numerous steady-state computations. In figure~\ref{impsing} and figure~\ref{imphist}, we show some of our transient-state results where the singular lid data (\ref{bcs}) in fact cause mesh divergence.  

\begin{figure}[ht] \centering
  {\includegraphics[keepaspectratio,height=4.5cm,width=16cm]{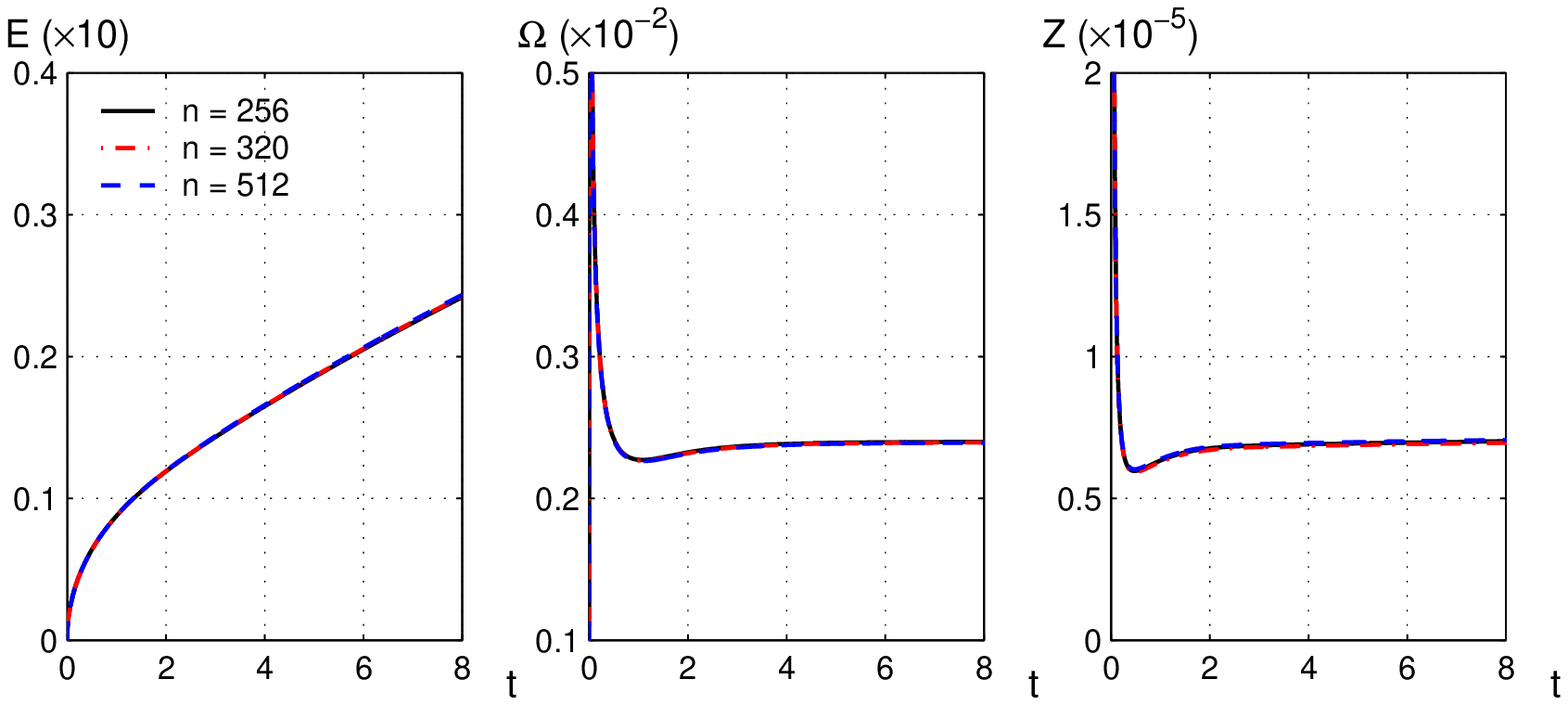}} 
 \caption{The ramp profile (\ref{ramp}) exhibits satisfactory mesh convergence. } \label{rmphist} 
\end{figure}

\begin{figure}[ht] \centering
  {\includegraphics[keepaspectratio,height=6cm,width=16cm]{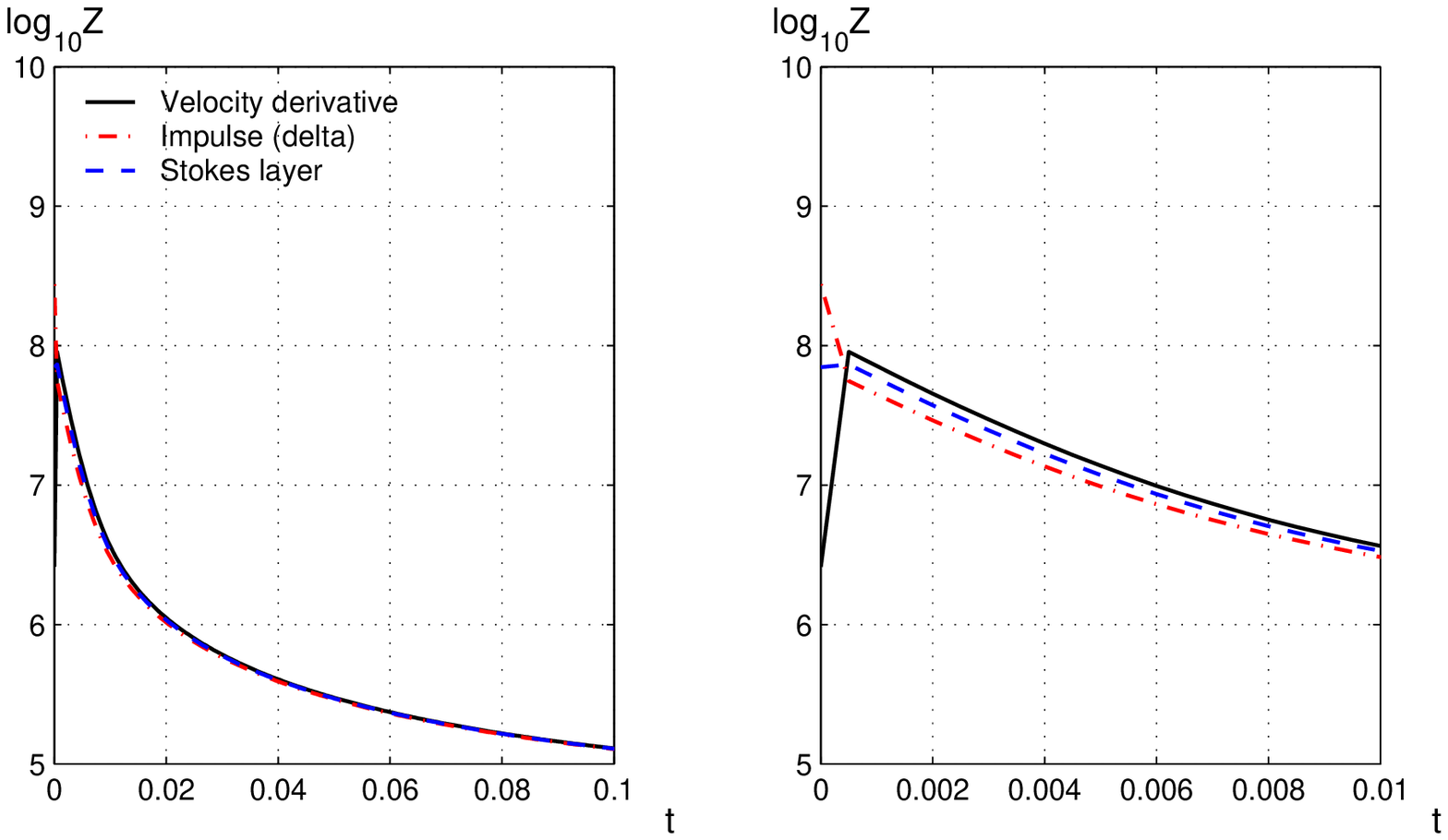}} 
 \caption{Palinstrophy overlay from three initial vorticity distributions with the same lid velocity profile (\ref{ramp}). The velocity derivative option refers to the initial wall $\zeta_0$ calculated from the continuity; the delta and Stokes data are given in (\ref{imps2}) and (\ref{imps}) respectively. Computations are carried out at $Re=1000$, $n=320$ and $\Delta t=5{\times}10^{-4}$. The current settling time $t^* \approx 0.03$.} \label{rmpzt0} 
\end{figure}

As the actual physics near the lid is a matter of complexity, the irregular lid velocity (\ref{bcs}) should be avoided altogether unless the singular corner flow is shown to be regular up to the third derivatives. We believe that the most promising method to move forward is to exploit regular lid velocity. Figure~\ref{rmphist} shows the good mesh convergence of the regulated ramp data (\ref{ramp}) (with $\zeta_0$ estimated from the continuity). In view of the mesh consistency, we present the calculations in figure~\ref{rmpzt0} that confirm the existence of a settling time (denoted by $t^*$), beyond which the flow is indistinguishable from different (non-zero) starting shears. The rate of the wall-shear production must strictly follow equation (\ref{vort}) in accordance with the given wall velocity. Even though we do not the have precise vorticity distribution on the lid at the beginning, the vorticity dynamics designates the compatible wall shear. In this respect, our iterative procedures are very effective to establish the compatibility solution. We may look at the temporal indefiniteness from a different perspective. Recall that the initialisation (\ref{ics}) is idealised. Unless lid's initial inertia is magically well-matched, the initial velocity is far from well-defined. What happens next is that the starting vorticity is modified immediately after the start, while the compatible vorticity field is dynamically established over an epoch interval so that the lid velocity profile attains the prescribed distribution. The end-product is the formation of the equilibrium wall viscous layer.


\begin{figure}[t] \centering
  {\includegraphics[keepaspectratio,height=17cm,width=17cm]{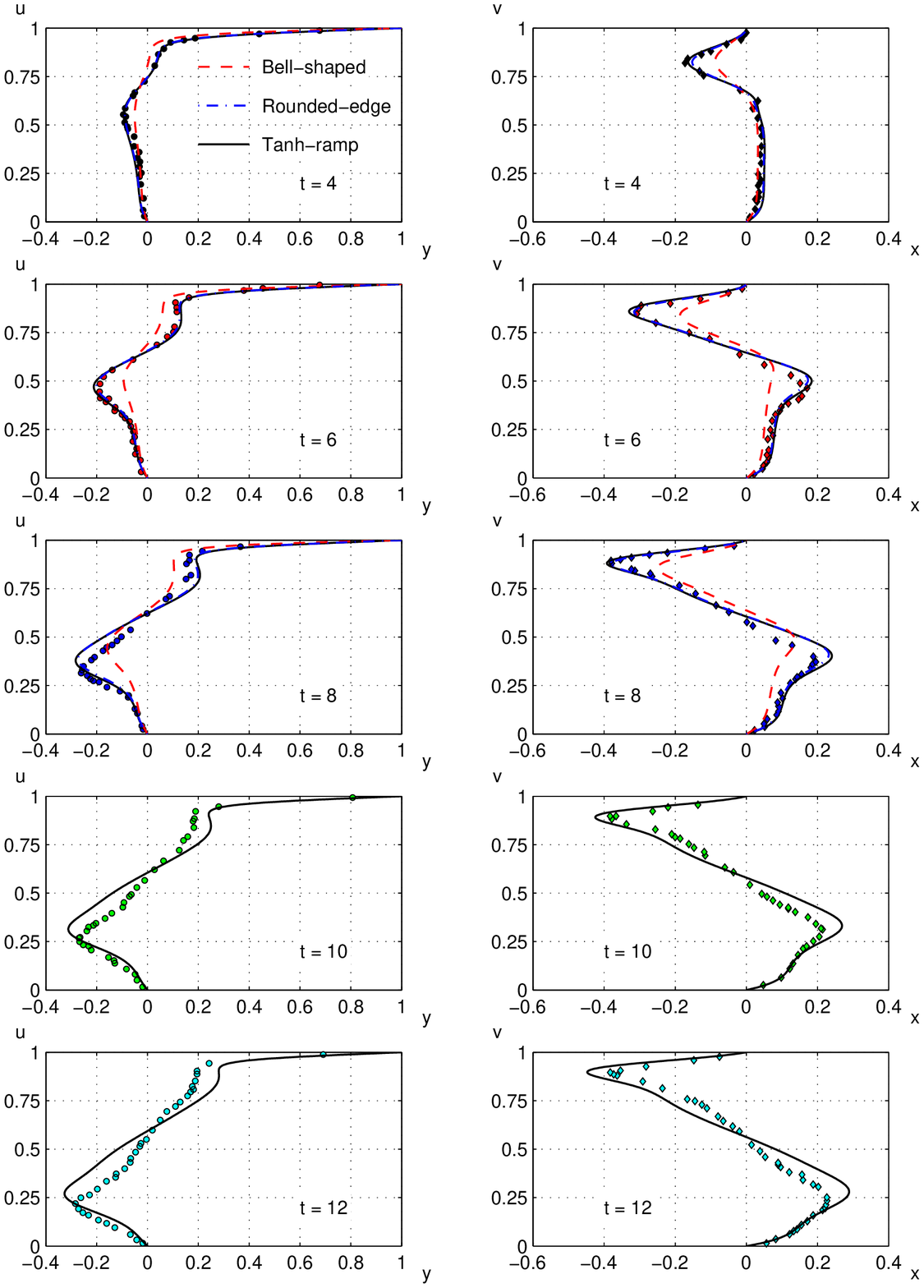}} 
 \caption{Computation results with the proposed lid conditions ($Re=1000$). The bell-shaped velocity predicts the experiment unsatisfactorily. In the rounded form and the tanh-ramp profile, there is a remarkable agreement with the tests of Guermond {\it et al}. In particular, the velocity slopes near the solid boundaries are well-predicted. At the later time $t>8$, the discrepancy in the regions away from the walls is largely due to three dimensional effects. Note that no attempts have been made to ``numerically modify'' the lid's acceleration in any part of our simulation. } \label{rmpexpt} 
\end{figure}

It is well-known that experiments on lid-driven cavity flow are delicate in practice. For visualisation purposes, well-made flow-fields at Reynolds numbers of a few thousands are difficult to realise, as pointed out by Guermond {\it et al}. (2002). To validate the present algorithm, comparison with tests is made in figure~\ref{rmpexpt}. Good agreement is found before the flow becomes truly $3$-dimensional. The key reason is that, because of the no-slip and the geometric symmetry, the flows close to the walls and in the mid-plane must be essentially $2$-dimensional during the early development. 

\begin{figure}[ht] \centering
  {\includegraphics[keepaspectratio,height=5.5cm,width=16cm]{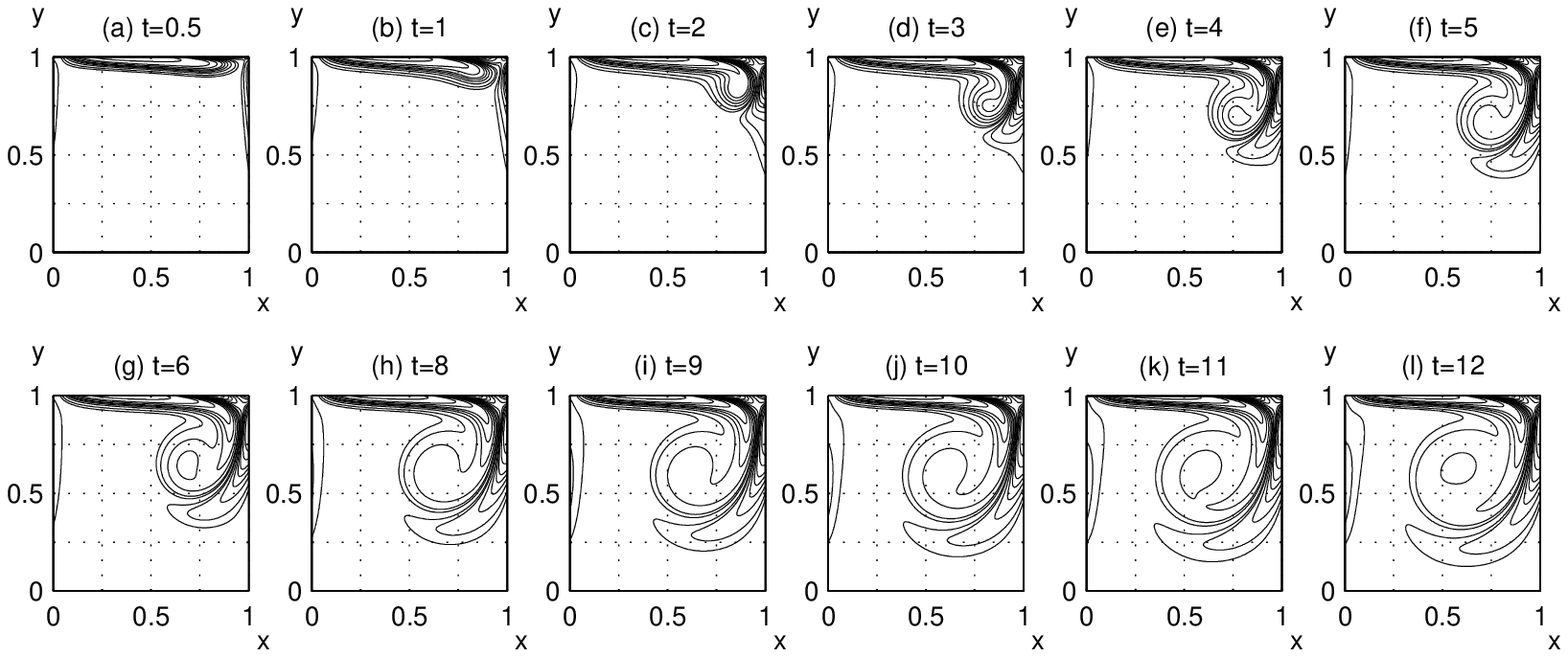}} 
 \caption{The initial transients of bell lid cavity (\ref{bell}) at $Re=1000$ and $n=256$. Plotted iso-vorticity contours are $\pm200, \pm100$, $\pm50$, $\pm20, \pm10, \pm7$, $\pm5, \pm4$, $\pm3, \pm2, \pm1$ and $\pm0.5$. } \label{bellsst} 
\end{figure}

\begin{figure}[ht] \centering
  {\includegraphics[keepaspectratio,height=5.5cm,width=16cm]{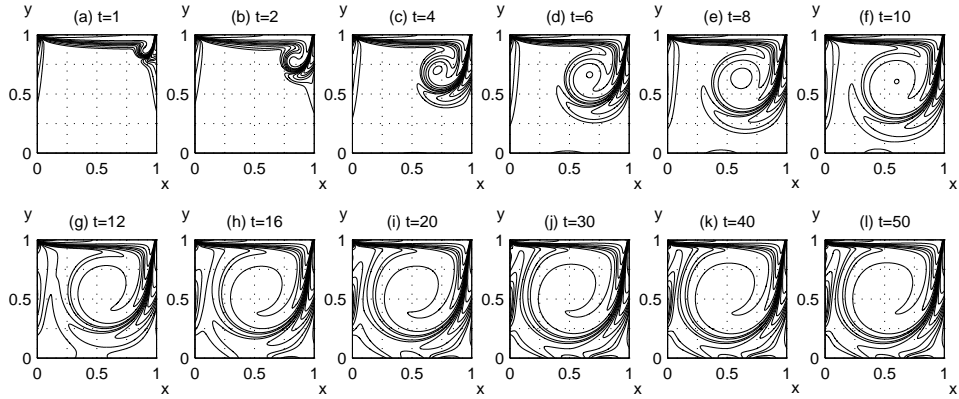}} 
 \caption{Navier-Stokes solutions of lid-driven cavity with ramp lid (\ref{ramp}), $Re=1000$, $n=512$, shown at the contours of the preceding plot. At $t=50$, the solutions almost reach the steady-state. The experimental data of Guermond {\it et al}. indicate that the initial flow settled to a quasi-stationary state at $t \approx 12$, and the start-up phase finalised at $t \approx 18$. Even in the test, the lid shear near the start is uncertain. Nevertheless, these simulation snap-shots are broadly in agreement with the observation. We can identify how the main vortex is developed from the lid, and the way it grows and migrates into the cavity from the upper right corner. In addition, the appearance of the wall viscous layers is well-explained.} \label{rmpsst} 
\end{figure}

\begin{figure}[ht] \centering
  {\includegraphics[keepaspectratio,height=17.5cm,width=17.5cm]{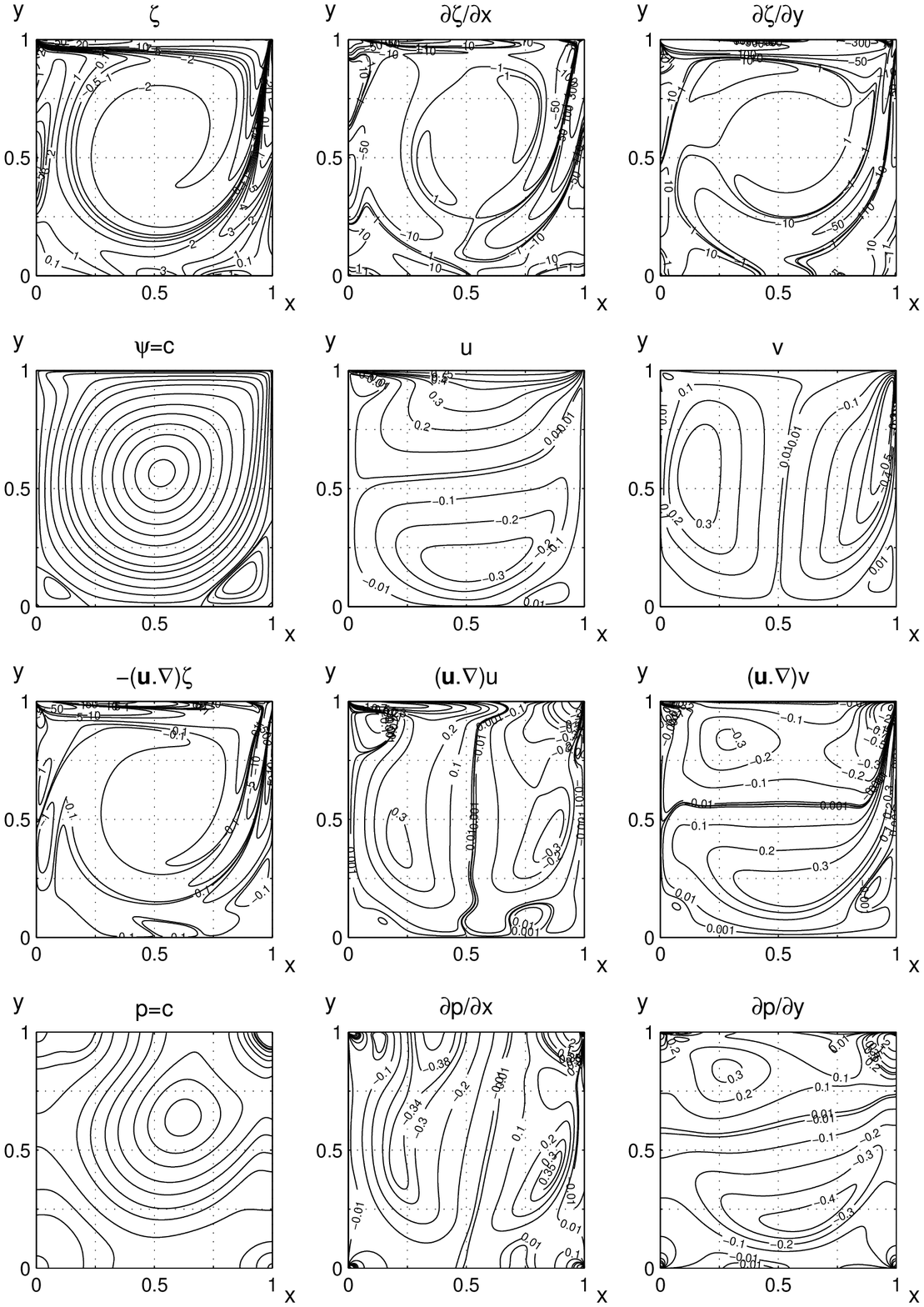}} 
 \caption{Selected flow solutions at $t=50$ (cf. figure~\ref{rmpsst}). } \label{rmp1k50p0} 
\end{figure}

The bell lid velocity is known to over-regulate the lid with weak vorticity production at the upper right corner at ($Re=1000$), as shown in figure~\ref{bellsst}. On the other hand, we are confident that the ramp distribution is well-suited in modelling the transients, see figure~\ref{rmpsst} for the evolution up to an effective steady-state. Specific numerics are highlighted in figure~\ref{rmp1k50p0}. The fact is that figures \ref{bellsst} and \ref{rmpsst} show dissimilar dynamics because of different $u_{\rm{Lid}}$, as the single dimensionless parameter $Re$ is unable to completely quantify the initial and boundary conditions. Experience shows an extremely long period of time is required for low-$\nu$ flows to attain a steady-state, assuming it exists. In many applications, there may not exist well-defined characteristic velocities, or they are unsteady and depend on viscosity over the transients. Since every flow has a beginning in principle as well as in practice, interpretations of the finite-time evolution in terms of Reynolds number seems to be over-simplistic.

\begin{figure}[t] \centering
  {\includegraphics[keepaspectratio,height=14.25cm,width=14cm]{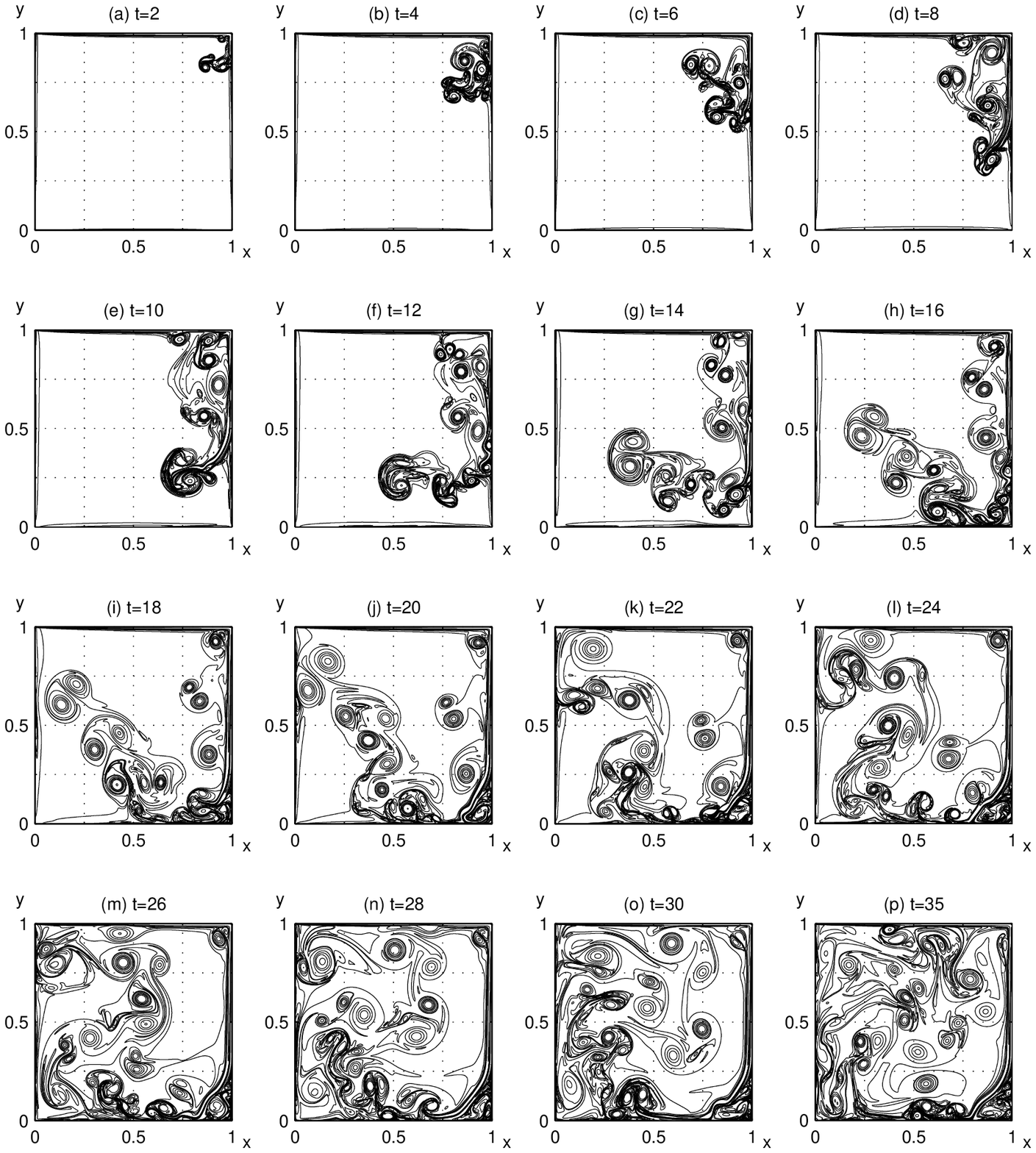}}
 \caption{Flow transients at $\nu=10^{-5}\: (n=2048, \Delta t=2.5{\times}10^{-4})$. Plotted iso-vorticity contours are $\pm150,\pm25$,$\pm10,\pm7$,$\pm5,\pm3,\pm1$ and $\pm0.1$. Concentrated eddies of small sizes are found soon after the start. They are either borne out of the wall shears, or strained from the periphery of the larger eddies. Every instantaneous field is full of profuse patches of mushroom vortices. The emergence of the small-scale vortices is an incessant event over time. Apart from the regions of the vortex cores and the walls, the remaining shear field contains weaker shears, and is less space-filling. To demarcate any vorticity-deficient zone is hard, because of the aggregate transport of the distant and neighbouring convoluted eddies. Up to the final time of the simulation ($t=50$), the viscous dissipation is rather mild; the hovering vortices roll up into isolated circular or elliptic shape. As these segregated eddies push and pull against one another, the velocity field must appear to be chaotic on observation. It is evident that the unsteady flow is neither homogeneous nor isotropic. The state of the evolution exemplifies genuine turbulence.} \label{rmp100k} 
\end{figure}

The convincing comparison with the experiments states that even the $2D$ theory has its relevance in the description of fluid motion in numerous circumstances. To further demonstrate the robustness of our numerical scheme, we have undertaken specific Navier-Stokes simulations at low viscosity with well resolved meshes. After a series of trial-and-error runs with different grid sizes and marching time $\Delta t$, we find that, within the present hardware capability, routine runs at grid $2048^2$ are feasible in practice. Figure~\ref{rmp100k} summarises one of our computations. 
\section{Conclusion}
We have shown that fluid dynamics in two dimensions is well-posed in the vorticity-stream function setting. In particular, the unique Dirichlet conditions $\psi_{\bdy}$ are the only boundary data needed to invert the $\psi$ Laplacian. A robust and efficient numerical method has been developed and validated against available experiments. On the basis of the present calculation, we see that the numerical solutions are only meaningful with proper examinations of mesh convergence where the palinstrophy, not the energy or the enstrophy, plays the crucial role.

For the problem of the lid-driven cavity, the (non-zero) vorticity at the start $t=0$ is immaterial; the dynamics establishes the equilibrium flow over the settling time.
Our computations at small viscosity $\nu \sim O(10^{-5})$ clearly show that turbulence ensues soon after the start, typically in time $t \sim O(10)$. One of the essential findings is that the viscous wall layers separate into the vortical structures with asymmetric vorticity derivatives, which further develop into mushroom-like vortices in large numbers. These eddies become more convoluted and eventually roll up into isolated vortices. Turbulent flows are a sea of interacting vortices of various sizes. 

At high Reynolds numbers, more refined meshes are required in order to capture the small-scale dynamics. Then computation is a matter of access to powerful machines. The essence is not whether we can do direct numerical simulations at grid $4096^3$ or $16384^3$ in a periodic box so as to establish a hierarchy of statistics quantities in the Fourier space. In the end, we must not forget that, in face of the myriad data generated by the brutal numerical force,  the Navier-Stokes dynamics is deterministic, and no real turbulent flows are homogeneous, isotropic or stationary.

\vspace{12mm}
\begin{acknowledgements}
\noindent 
11 April 2018

\noindent 
\texttt{f.lam11@yahoo.com}
\end{acknowledgements}
\end{document}